\documentclass[twocolumn,superscriptaddress,prl]{revtex4-1}
\usepackage{amsmath}
\usepackage{amssymb}
\usepackage{graphicx}
\usepackage{esint}
\usepackage{epstopdf}
\usepackage{amsfonts}
\usepackage{tabularx}
\usepackage{dcolumn}
\usepackage{bm}
\usepackage[unicode=true,pdfusetitle,pdfborder={0 0 1}]{hyperref}
\hypersetup{colorlinks,
linkcolor=blue,          citecolor=blue,        filecolor=blue,      urlcolor=blue           }

\makeatletter

\@ifundefined{textcolor}{}
{
 \definecolor{BLACK}{gray}{0}
 \definecolor{WHITE}{gray}{1}
 \definecolor{RED}{rgb}{1,0,0}
 \definecolor{GREEN}{rgb}{0,1,0}
 \definecolor{BLUE}{rgb}{0,0,1}
 \definecolor{CYAN}{cmyk}{1,0,0,0}
 \definecolor{MAGENTA}{cmyk}{0,1,0,0}
 \definecolor{YELLOW}{cmyk}{0,0,1,0}
}

\makeatother

\begin{document}

\title{Probe knots and Hopf insulators with ultracold atoms}

\author{Dong-Ling Deng}
\affiliation{Department of Physics, University of Michigan, Ann Arbor, Michigan 48109, USA}
\affiliation{Condensed Matter Theory Center and Joint Quantum Institute, Department
of Physics, University of Maryland, College Park, MD 20742-4111, USA}
\affiliation{Center for Quantum Information, IIIS, Tsinghua University, Beijing 100084,
PR China}

\author{Sheng-Tao Wang}
\affiliation{Department of Physics, University of Michigan, Ann Arbor, Michigan 48109, USA}
\affiliation{Center for Quantum Information, IIIS, Tsinghua University, Beijing 100084,
PR China}

\author{Kai Sun}
\affiliation{Department of Physics, University of Michigan, Ann Arbor, Michigan 48109, USA}

\author{L.-M. Duan}
\affiliation{Department of Physics, University of Michigan, Ann Arbor, Michigan 48109, USA}
\affiliation{Center for Quantum Information, IIIS, Tsinghua University, Beijing 100084,
PR China}

\begin{abstract}
\textbf{Knots and links are fascinating and intricate topological objects.
Their influence spans from DNA and molecular chemistry to vortices in superfluid helium, defects in liquid crystals and cosmic strings in the early
universe. Here, we find that knotted structures also exist in a peculiar class of three-dimensional topological insulators---the Hopf insulators.
In particular, we demonstrate that the momentum-space spin textures of Hopf insulators are twisted in a nontrivial way, which implies the presence of various
knot and link structures. We further illustrate that the knots and nontrivial spin textures can be probed via standard time-of-flight images in cold atoms as preimage contours of spin orientations in stereographic coordinates. The extracted Hopf invariants, knots, and links are validated to be robust to typical experimental imperfections. Our work establishes the existence of knotted structures in Hopf insulators, which may have potential applications in spintronics and quantum information processing. }
\end{abstract}

\maketitle

\noindent
More than a century ago, Lord Kelvin propounded his celebrated ``vortex
atom'' theory, which  asserted that atoms are made of vortex knots in the
aether \cite{kelvin1867vortex}. Knot theory has since become
a central subject in topology and began to undertake important roles in diverse fields of sciences. Biologists discovered that molecular knots and links in DNA are crucial in vital processes of replication, transcription
and recombination \cite{bates2005dna,meluzzi2010biophysics}. In supramolecular
chemistry, complex knotted structures have been demonstrated in the
laboratory \cite{chichak2004molecular, han2010folding, ponnuswamy2012discovery}
and were shown to be essential for the crystallization and rheological
properties of polymers \cite{Gennes1979Scaling}. In physics, knot has appeared in many subfields,
such as classical field theory \cite{faddeev1997stable,Kedia2013Tying}, helium
superfluid \cite{ricca1999evolution}, fluid mechanics and plasma
\cite{woltjer1958theorem}, spinor Bose-Einstein condensates \cite{kawaguchi2008knots}, chiral nematic colloids \cite{machon2013knots,Machon2014knotted,Alexander2012Colloquium}, quantum chromodynamics \cite{buniy2003model} and string theory \cite{witten1989quantum}. Recently, exciting progresses are made in laboratories to observe the knots. Experimental creation of knotted configurations
has been demonstrated in liquid crystal \cite{senyuk2013topological, tkalec2011reconfigurable, Chen2013generating, martinez2014mutually}, laser light \cite{dennis2010isolated}, fluid flows \cite{kleckner2013creation, irvine2014liquid}, and spinor Bose-Einstein condensate \cite{Hall2016Tying}, driving a new wave of interest in the study of knots.

In quantum condensed matter physics, topology is crucial in understanding exotic phases and phase transitions. Notable examples include the discoveries of the Berezinskii-Kosterlitz-Thouless transition \cite{Berezinskii1971Destruction,Kosterlitz1973Ordering} and  the quantum Hall effect \cite{Cage2012Quantum}.  More recently,
another new class of materials, dubbed topological insulators, was theoretically predicted and experimentally observed \cite{qi2011topological, hasan2010colloquium, moore2010birth}. In general, they are topological phases protected by system symmetries, which cannot be smoothly connected to the trivial phase if the respective symmetries are preserved \cite{chen2012symmetry}. For these materials,
the nontriviality originates from the interplay between symmetry and topology. Many peculiar physical phenomena are predicted or observed in experiments, including, for instance, quantized Hall conductance \cite{thouless1982quantized}, robust chiral edge states \cite{qi2011topological}, magnetic monopole \cite{qi2009inducing}, wormhole  and Witten  effects \cite{rosenberg2010wormhole,rosenberg2010witten}. However,  despite these striking progresses, the relation between topological insulators and knot theory remains largely unexplored. It is unclear whether topological insulators carry nontrivial knot structures and how we can visualize these structures. 



In this paper, we reveal intriguing knot and link structures hidden in a particular class of topological insulators. Specifically, we find various nontrivial knot and link structures encoded in the spin textures of Hopf insulators. Hopf insulators are special three-dimensional (3D) topological insulators without the need for symmetry protection \cite{moore2008topological, deng2013hopf}. We show that in momentum space the spin textures of Hopf insulators represent  special realizations of the long-sought-after Hopfions \cite{faddeev1975quantization, faddeev1997stable}, which are 3D topological solitons evading any experimental observation so far. Inspired by the rapid experimental progress of synthetic gauge fields in cold atoms \cite{lin2011spin, galitski2013spin, beeler2013spin, dalibard2011colloquium, bloch2012quantum,Goldman2016Topological,Jotzu2014Experimental,Wu2015Realization},
we propose a scheme to measure the Hopf topology and visualize different knots and links based on cold-atom technologies. We demonstrate how to extract the Hopf invariant and the knotted spin textures from the state-of-the-art time-of-flight imaging data. The corresponding knots and links are explicitly revealed as preimage contours of the observed spin orientations via a stereographic coordinate system. The extracted information---Hopf invariants, knots, and links---is robust against typical experimental imperfections. This work opens a new avenue for study of topological insulators and knot theory in ultracold atom experiments.  \newline

\noindent  \textbf{\large{Results} }

\noindent  \textbf{Hopf insulators and knots.} Hopf insulators are 3D
topological insulators characterized by an integer Hopf index and have topologically protected metallic surface states \cite{moore2008topological, deng2013hopf}. Unlike the recently predicted \cite{fu2007topological, moore2007topological, roy2009topological}
and experimentally observed 3D $\mathbb{Z}_{2}$ topological insulators with a time-reversal symmetry
\cite{hsieh2008topological,xia2009observation}, Hopf insulators do
not require any symmetry protection (other than the prerequisite $U(1)$ charge conservation). They are
peculiar exceptions that sit outside of the periodic table for topological insulators and superconductors \cite{schnyder2008classification, kitaev2009periodic}. One may regard them as a 3D generalization of the 2D quantum anomalous Hall effect \cite{chang2013experimental}. A model Hamiltonian for the Hopf insulator with Hopf invariant $\chi=\pm1$ was first introduced by Moore, Ran and Wen \cite{moore2008topological}. In
our previous works \cite{deng2013hopf, deng2014systematic}, we generalized
their results and constructed Hamiltonians for Hopf insulators
with arbitrary Hopf index, using two different approaches, one
based on the quaternion algebra \cite{deng2014systematic} and the
other based on the generalized Hopf map \cite{deng2013hopf} $f:\;\mathbb{S}^{3} \rightarrow \mathbb{S}^{2}$ (up to irrelevant overall normalizations)
\begin{equation}
S_{x}+i S_{y}=2\eta_{\uparrow}^{p}\bar{\eta}_{\downarrow}^{q},\; S_{z}= (|\eta_{\uparrow}|^{2p}-|\eta_{\downarrow}|^{2q}), \label{eq:HopfMap}
\end{equation}
where $\mathbb{S}^2$ and $\mathbb{S}^3$ denote respectively the 2D and 3D spheres , the  $\mathbb{S}^3$ coordinates  $\boldsymbol{\eta}=(\text{Re}[\eta_{\uparrow}],\text{Im}[\eta_{\uparrow}],\text{Re}[\eta_{\downarrow}],\text{Im}[\eta_{\downarrow}])$ are mapped to $\mathbb{S}^2$ coordinates $(S_{x},S_{y},S_{z})$, and $p$, $q$ are integers prime to each other. The Hopf invariant for the map $f$ is known to be $\chi(f)=\pm pq$
with the sign determined by the orientation of the three-sphere \cite{whitehead1947expression}. To construct physical Hamiltonians, $\boldsymbol{\eta}$ can in turn be considered as another map $g$ from the first
Brillouin zone (BZ) to the three-sphere: $\mathbb{T}^{3}\rightarrow\mathbb{S}^{3}$
\begin{align}
\eta_{\uparrow}\mathbf{(k)} & =\sin k_{x}+i\sin k_{y}, \notag \\
\eta_{\downarrow}\mathbf{(k)}&=\sin k_{z}+i(\cos k_{x}+\cos k_{y}+\cos k_{z}+h), \label{eq:map_g}
\end{align}
where $\mathbb{T}^3$ is a 3D torus (describing the first BZ).
Thus, the map $\mathbf{S(k)}=\left( S_{x}(\mathbf{k}),S_{y}(\mathbf{k}),S_{z}(\mathbf{k}) \right)$ can be regarded as a composition of two maps, $\mathbf{S}=f \circ g$. In momentum space, we construct a tight-binding Hamiltonian  based on $\mathbf{S(k)}$ through  $H=\sum_{\mathbf{k}}\psi_{\mathbf{k}}^{\dagger}\mathcal{H}(\mathbf{k})\psi_{\mathbf{k}}$ with $\psi_{\mathbf{k}}^{\dagger}=(c_{\mathbf{k}\uparrow}^{\dagger},c_{\mathbf{k}\downarrow}^{\dagger})$
and
\begin{equation}
\mathcal{H}(\mathbf{k})  =  \mathbf{S}(\mathbf{k})\cdot\boldsymbol{\sigma},\label{eq:Hopf Ham-MomentumS}
\end{equation}
where $c_{\mathbf{k}\uparrow}^{\dagger},c_{\mathbf{k}\downarrow}^{\dagger}$ are fermionic creation operators and $\boldsymbol{\sigma}=(\sigma^{x},\sigma^{y},\sigma^{z})$ are three Pauli matrices. The Hamiltonian $\mathcal{H}(\mathbf{k})$ contains $\left(p+q\right)$th order polynomials of $\sin(\mathbf{k})$ and $\cos(\mathbf{k})$, which correspond to $\left(p+q\right)$-th neighbor hoppings in real space \cite{deng2013hopf}.
\begin{figure}[!t]
\includegraphics[width=0.49\textwidth]{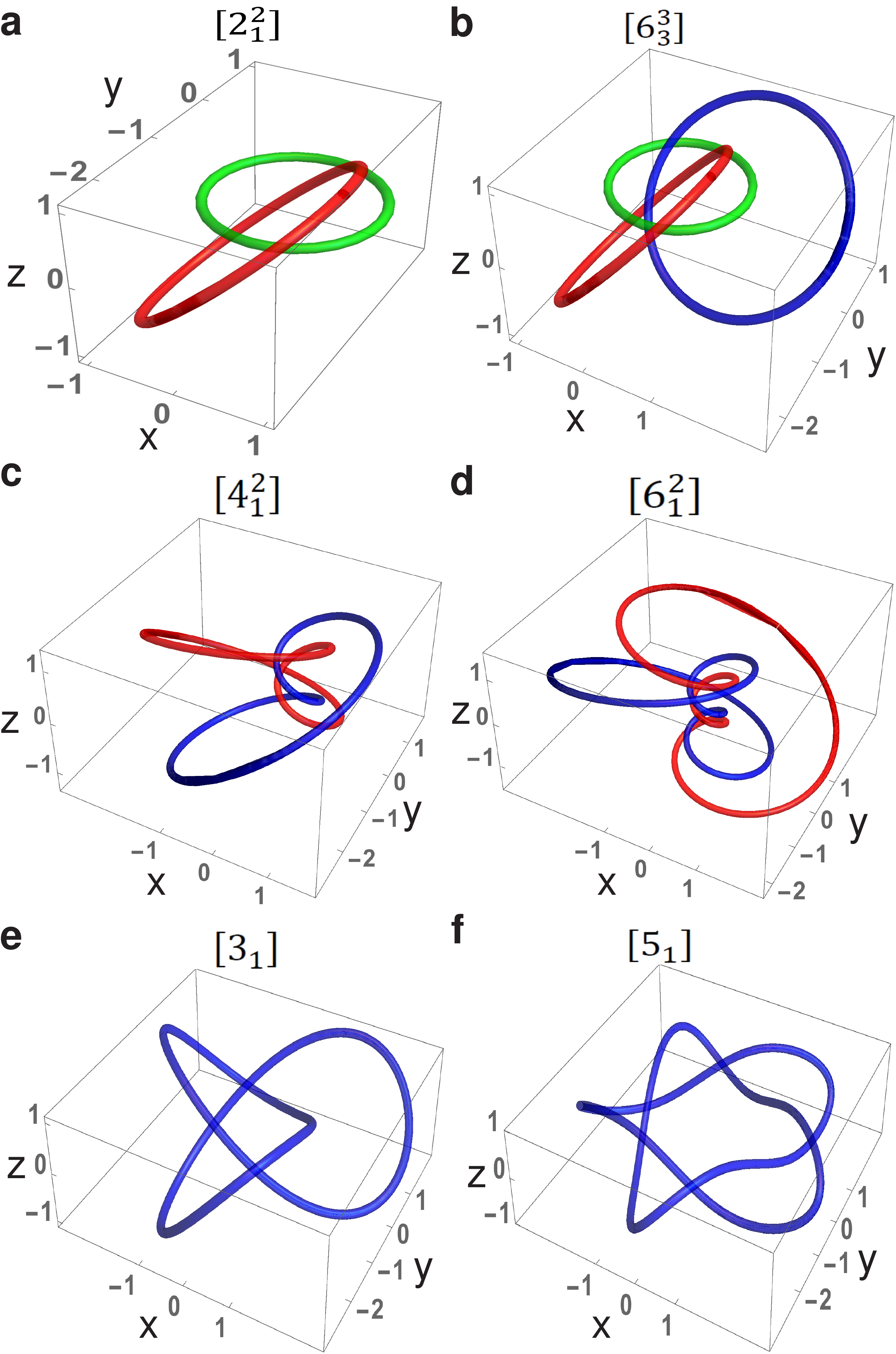}
\caption{\textbf{Knots and links hidden in Hopf insulators.} These knots and links correspond to preimages of three different spin orientations $\hat{\mathbf{S}}_{1}=(1,0,0)$, $\hat{\mathbf{S}}_{2}=(0,1,0)$ and $\hat{\mathbf{S}}_{3}=(0,0,1)$ living on $\mathbb{S}^{2}$. The loops in red (blue, green) are preimages of $\hat{\mathbf{S}}_{1}$
($\hat{\mathbf{S}}_{2}$, $\hat{\mathbf{S}}_{3}$) obtained
from the contour plot $f^{-1}(\hat{\mathbf{S}}_{1})$ $\big( f^{-1}(\hat{\mathbf{S}}_{2}), f^{-1}(\hat{\mathbf{S}}_{3}) \big)$ in a stereographic coordinate system (see Methods section). The symbols inside each square bracket denote the standard
Alexander\textendash{}Briggs notation for the corresponding knot (link).
The four links are:  (\textbf{a}) the Hopf link,  (\textbf{b}) the $6_{3}^{3}$ link,  (\textbf{c}) the Solomon's link, and (\textbf{d}) the $6_{1}^{2}$ link. The two knots are: (\textbf{e}) the trefoil knot
and  (\textbf{f}) the Solomon seal knot. The parameters are chosen
as: $p=q=1$ in (\textbf{a}) and (\textbf{b}); $p=1,\; q=2$ in (\textbf{c});
$p=1,\; q=3$ in (\textbf{d}); $p=3,\; q=2$ in (\textbf{e}) and $p=5,\; q=2$
in (\textbf{f}). $h=2$ for (\textbf{a}-\textbf{f}).  From (\textbf{a}-\textbf{d}), it is evident that
the linking number of two different preimage contours is equal to
the Hopf invariant $\chi(f)$. \label{Hopflink}}
\end{figure}

To study the topological properties of the Hamiltonian in Eq.\ (\ref{eq:Hopf Ham-MomentumS}), we define a normalized (pseudo-) spin field $\hat{\mathbf{S}}(\mathbf{k})=\mathbf{S}(\mathbf{k})/|\mathbf{S}(\mathbf{k})|$ (the normalization does not affect the topological properties). Hopf invariant, also known as Hopf charge or Hopf index, can be computed as an integral \cite{moore2008topological, deng2013hopf}:
\begin{equation}
\chi(\hat{\mathbf{S}})=-\int_{\text{BZ}}\mathbf{F\cdot\mathbf{A}}\; d^{3}\mathbf{k,}\label{eq:Hopf-Index-1}
\end{equation}
where $\mathbf{F}$ is the Berry curvature defined as $F_{\mu}=\frac{1}{8\pi}\epsilon_{\mu\nu\tau}\hat{\mathbf{S}}\cdot(\partial_{\nu}\mathbf{\hat{\mathbf{S}}\times\partial_{\tau}\hat{\mathbf{S}}})$ with $\epsilon_{\mu\nu\tau}$ being the Levi-Civita symbol and $\partial_{\nu, \tau} \equiv \partial_{k_{\nu,\tau}}$ ($\mu, \nu, \tau \in \{ x,y,z \}$), and $\mathbf{A}$ is the associated Berry connection satisfying $\nabla\times\mathbf{A}=\mathbf{F}$. Direct calculations give $\chi(\hat{\mathbf{S}})=\pm pq$ if $1<|h|<3$, $\chi(\hat{\mathbf{S}})=\pm2pq$ if $|h|<1$, and $\chi(\hat{\mathbf{S}})=0$ otherwise \cite{deng2013hopf}. This can be understood intuitively by decomposing the composition map, $\chi(\hat{\mathbf{S}}) = \chi(f)\Lambda(g)=\pm pq\Lambda(g)$, where the map $g$ is classified by another topological invariant
\begin{equation}
\Lambda(g)  =  \frac{1}{12\pi^{2}}\int_{\text{BZ}}d\mathbf{k}\epsilon_{\mu\nu\rho\tau}\frac{\epsilon_{\alpha\beta\gamma}}{|\boldsymbol{\eta}|^{4}}\boldsymbol{\eta}_{\mu}\partial_{\alpha}\boldsymbol{\eta}_{\nu}\partial_{\beta}\boldsymbol{\eta}_{\rho}\partial_{\gamma}\boldsymbol{\eta}_{\tau}.\label{eq:GammaIndex}
\end{equation}
With $g$ given by Eq.\ (\ref{eq:map_g}), one
obtains $\Lambda(g)=1$ if $1<|h|<3$, $\Lambda(g)=-2$ if $|h|<1$,
and $\Lambda(g)=0$ otherwise. Geometrically, $\Lambda(g)$ counts how many times $\mathbb{T}^{3}$ wraps around $\mathbb{S}^{3}$ nontrivially under the map $g,$ and $\chi(f)$ describes how many times $\mathbb{S}^{3}$ wraps around $\mathbb{S}^{2}$ nontrivially under $f$. Their composition gives the Hopf invariant $\chi(\hat{\mathbf{S}})$ \cite{deng2013hopf}.

The spin field $\hat{\mathbf{S}}$ can be viewed as a map from $\mathbb{T}^{3}$ to $\mathbb{S}^{2}$. While the domain $\mathbb{T}^{3}$ is three-dimensional, the target space $\mathbb{S}^{2}$ is two-dimensional. As a consequence, the preimage of a point in $\mathbb{S}^{2}$ should be a closed loop in $\mathbb{T}^{3}$, and the linking number of two such loops corresponds to the Hopf invariant $\chi(\hat{\mathbf{S}})$. Similarly, the linking number of two preimage loops in $\mathbb{S}^{3}$ under the map $f$ gives the Hopf invariant $\chi(f)$, which is part of $\chi(\hat{\mathbf{S}})$. For easy visualization of knots and links, we work with $\mathbb{S}^{3}$ rather than $\mathbb{T}^{3}$ and probe the Hopf index $\chi(f)$.  Fig.\ \ref{Hopflink} shows several links and knots corresponding to certain spin orientations by using stereographic coordinates to represent $\mathbb{S}^{3}$ (see Methods section). It is clear from the figure that the linking number of two preimage contours of distinct spin orientations is equal to the Hopf invariant $\chi(f)$. More interestingly, we note that even a single preimage
contour may form a highly nontrivial knot when $p$ and $q$ become large. For instance, the trefoil
knot and the Solomon seal knot plotted in Fig.\ \ref{Hopflink}\textbf{e}
and Fig.\ \ref{Hopflink}\textbf{f} are two well-known nontrivial knots with nonunit knot polynomials \cite{kauffman2013knots}. More complex knots and links emerge for larger $p$ and $q$.

\begin{figure}
\includegraphics[width=0.4\textwidth]{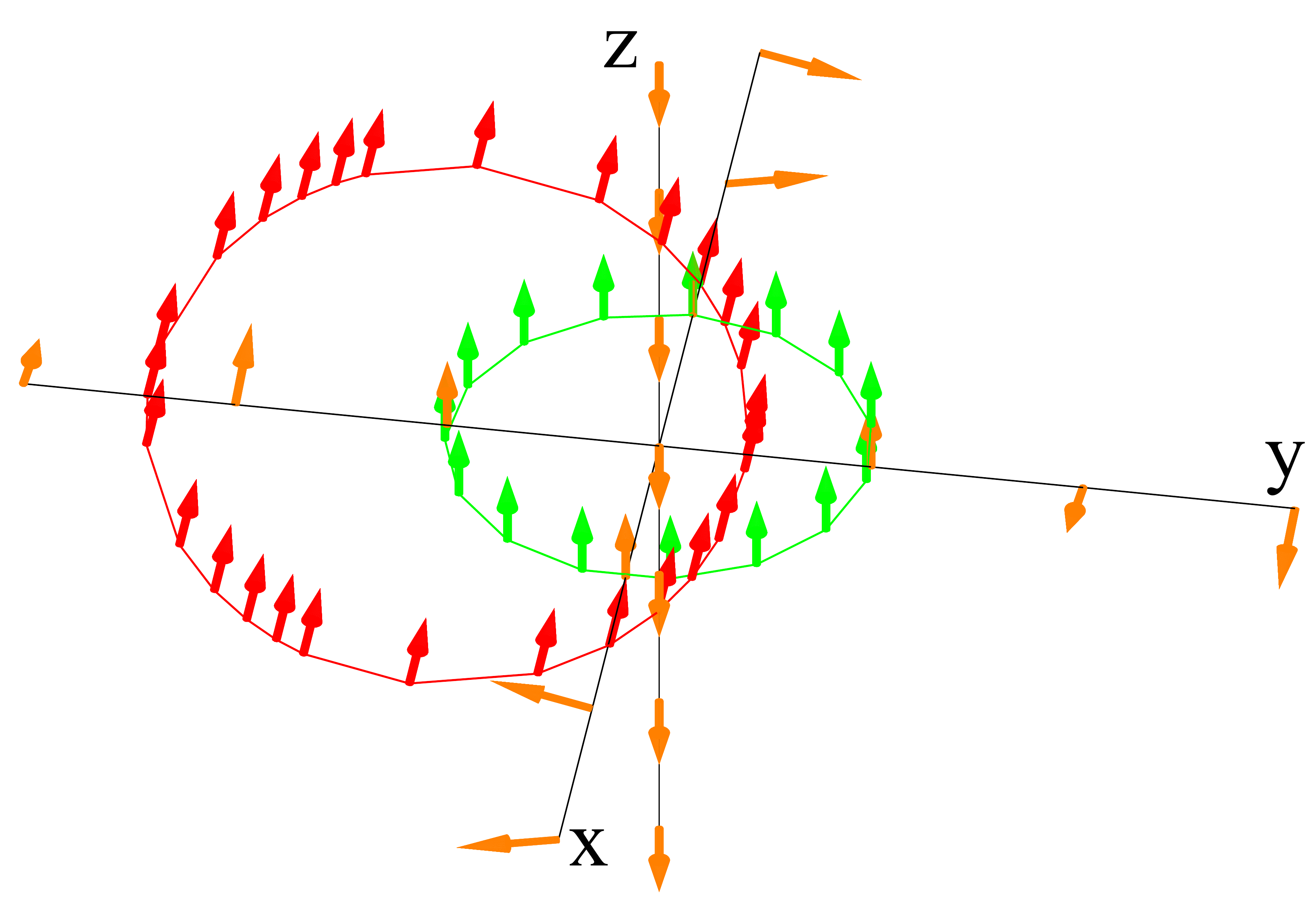}

\caption{\textbf{\small{Knotted spin texture in stereographic coordinates.}} We sketch out the spin orientations along each
axis and on two circles. The parameters are chosen as $p=q=1$ and
$h=2$. Spins reside on the red (green) circle point to the $x$ $(z)$
direction and those on the $z$ axis all point to the south (negative $z$ direction). Also, all spins faraway from the origin asymptotically point
to the south. This spin texture is nontrivial (twisted with $\chi(f)=1$)
and cannot be untwined continuously unless a topological phase transition
is crossed.\label{fig:Knotted-spin-texture}}
\end{figure}

A nonvanishing value of $\chi(\hat{\mathbf{S}})$ also indicates that
the spin field $\hat{\mathbf{S}}$ has a nontrivial texture that cannot
be continuously deformed into a trivial one. Mathematically, the Hopf
invariant in Eq.\ (\ref{eq:Hopf-Index-1}) is a characteristic topological
invariant of a sphere fiber bundle and a nonvanishing $\chi$ generally
precludes the existence of a global section due to the obstruction
theory, in analogy to the scenario where nonzero Chern numbers forbid the tangent bundle of a two-sphere to have a global section~\cite{hatcher2002algebraic}. A physical interpretation is that one can never untwist $\hat{\mathbf{S}}$ smoothly unless a topological phase transition is crossed. In Fig.~\ref{fig:Knotted-spin-texture}, the simplest nontrivial spin texture corresponding to $\chi(f)=1$ is sketched. One may regard $\hat{\mathbf{S}}$ as a unit continuous vector field. In the stereographic coordinates, $\hat{\mathbf{S}}$ resembles a nontrivial solution to the Faddeev-Skyrme model \cite{faddeev1975quantization} with Hopf charge one. In this sense, $\hat{\mathbf{S}}$ is a special realization of Hopfions, which are 3D topological solitons with broad applications but has hitherto escaped experimental observations \cite{hietarinta2012scattering}. In the next section, we will show that all these nontrivial knots, links, and spin textures can be measured in cold-atom experiments with time-of-flight imaging. \newline

\noindent \textbf{Probe knots and topology in optical lattices.} Hopf
insulators are special topological insulators that have not been
seen in solid systems. Although some magnetic compounds such
as $\text{R}_{2}\text{Mo}_{2}\text{O}_{7}$, with R being a rare earth
ion, are proposed to be possible candidates \cite{moore2008topological},
actual realization of the Hopf insulator phase in solid remains very challenging due to the complicated spin-orbit couplings. The rapid experimental progress in synthetic spin-orbit couplings with cold atoms \cite{lin2011spin, galitski2013spin, beeler2013spin, dalibard2011colloquium, bloch2012quantum,Goldman2016Topological,Jotzu2014Experimental,Wu2015Realization} provides a new promising platform to simulate various topological phases.
Actually, some model Hamiltonians, which are initially proposed mainly for theoretical studies and thought to be out of reach for experiments due to their unusual and complex couplings, have indeed been realized in cold-atom labs. For instance, an experimental observation of the topological Haldane model with ultracold atoms has been achieved recently \cite{Jotzu2014Experimental}. It is likely that Hopf insulators will also have a cold-atom experimental implementation in the near future. With this in mind, we illustrate below how to detect the Hopf invariant and probe the knot structure with ultracold atoms in optical lattices assuming a Hopf Hamiltonian is realized. In Ref.\ \cite{deng2014direct}, {\color{red} we} introduced a generic method to directly measure various topological
invariants based on time-of-flight imaging of cold atoms. This
method is applicable to the detection of topological band insulators in
any spatial dimensions. Here we apply it to the detection of Hopf insulators in cold-atom systems.

\begin{table}
\caption{\textbf{\small{Hopf index extracted from simulated time-of-flight images.}} Simulated experimental results for Hopf invariants with different lattice sizes and varying $h$. Four different situations, one with periodic and
three with open boundary conditions, are considered. The Hopf index
can be found from the momentum density distributions obtained
directly through time-of-flight imaging. The parameters used are $p=q=1$ and $\gamma_{\text{t}}=\gamma_{\text{r}}=0.1$.}

\begin{ruledtabular}%
\begin{tabular}{cccccc}
h & Size  & Periodic & Open  & Trap  & Pert.+Trap \tabularnewline
\hline  \vspace{-.3cm} \tabularnewline
0  & $10^{3}$  & -2.058 & -1.956 & -1.985  & -1.986\tabularnewline
0 & $20^{3}$  & -2.019 & -2.019 & -2.025  & -2.025\tabularnewline
2 & $10^{3}$  & 1.041 & 0.982 & 0.986 & 0.986\tabularnewline
2 & $20^{3}$  & 1.012 & 1.008 & 1.009  & 1.009\tabularnewline
4 & $20^{3}$  & $-9.6\times10^{-5}$ & $2.9\times10^{-5}$ & $6.6\times10^{-5}$ & $6.7\times10^{-5}$ \tabularnewline
\end{tabular}\end{ruledtabular} \label{Numerics-Table}
\end{table}

From Eq.\ (\ref{eq:Hopf-Index-1}), to extract the Hopf invariant,
it is essential to obtain the spin texture $\hat{\mathbf{S}}(\mathbf{k})$
in momentum space. In experiment, one discretizes the BZ and a
pixelized version of $\hat{\mathbf{S}}(\mathbf{k})$ can be obtained
through time-of-flight imaging. Refs \cite{alba2011seeing, deng2014direct} describe in detail how to measure $\hat{\mathbf{S}}(\mathbf{k})$ in experiment. One astute observation \cite{alba2011seeing} was that the spin component is related to the density distributions as $\hat{S}_{z}(\mathbf{k})=[n_{\uparrow}(\mathbf{k})-n_{\downarrow}(\mathbf{k})]/[n_{\uparrow}(\mathbf{k})+n_{\downarrow}(\mathbf{k})]$. A fast Raman or radio frequency pulse before time-of-flight rotates the atomic states and maps $\hat{S}_{x}$ and $\hat{S}_{y}$ to $\hat{S}_{z}$, enabling one to reconstruct the whole pixelized $\hat{\mathbf{S}}(\mathbf{k})$. To obtain the 3D momentum distributions, one may map out the 2D densities  $n(k_{x},k_{y},k_{z_{i}})$ with various $k_{z_{i}}$ layers \cite{deng2014direct}. It is encouraging to see in Table \ref{Numerics-Table} that ten or twenty layers are sufficient to produce very good results.

With the measured $\hat{\mathbf{S}}(\mathbf{k})$ in hand, we can extract the Hopf index directly. As $F_{\mu}=\frac{1}{8\pi}\epsilon_{\mu\nu\tau}\hat{\mathbf{S}}\cdot(\partial_{\nu}\mathbf{\hat{\mathbf{S}}\times\partial_{\tau}\hat{\mathbf{S}}})$,
we obtain the Berry curvature $\mathbf{F}$ at each pixel of the
BZ. We then find the Berry connection by solving a discrete version of the equation $\nabla\times\mathbf{A}=\mathbf{F}$ in the Coulomb gauge $\nabla\cdot\mathbf{A}=0$. Hopf index can thus be extracted from Eq.~(\ref{eq:Hopf-Index-1}), replacing the integral by a discrete summation. To simulate real experiments, we write down the Hamiltonian $H$ in real space and consider a finite-size lattice with open boundaries. In addition, we add two terms into the Hamiltonian to account for typical experimental imperfections. The first one is a global harmonic trap parametrized by $\gamma_{\text{t}}$, and the second one is a random noise characterized by $\gamma_{\text{r}}$ (see Methods section). For the simplest case of $p=q=1$, we numerically diagonalize the realistic real-space Hamiltonian and compute the corresponding Hopf index for different $h$ based on the method introduced in Ref.~\cite{deng2014direct}. Our results are summarized in Table \ref{Numerics-Table}. We can see that the Hopf
index converges rapidly to the expected value as the lattice size increases
and the detection method remains robust against typical experimental
imperfections.

\begin{figure}
\includegraphics[width=0.49\textwidth]{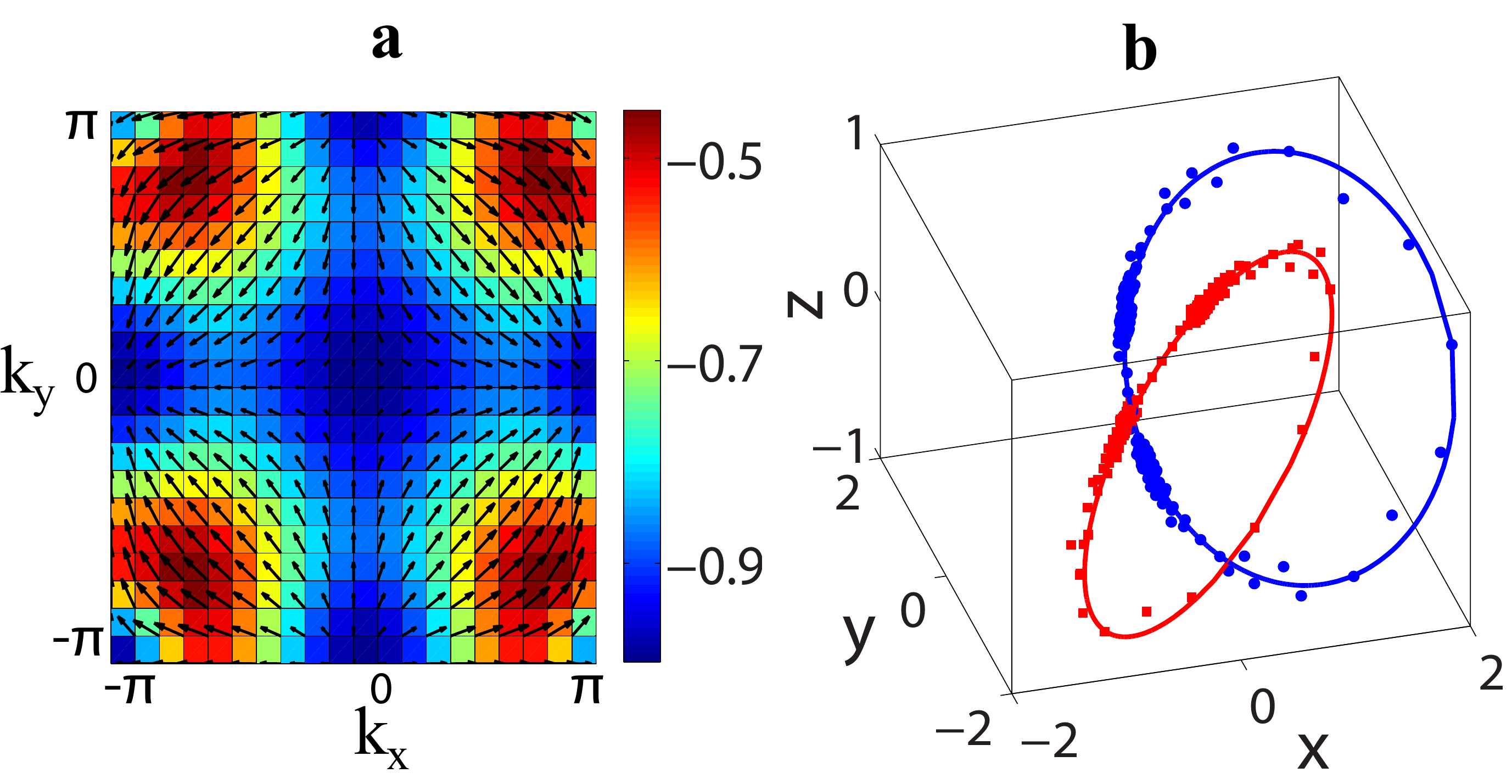}

\caption{\textbf{\small{Spin texture and the Hopf link from numerical simulations of real experiments.}} (\textbf{\small{a}})
Simulated spin texture in the $k_{x}$-$k_{y}$ plane with $k_{z}=0$.
The background color scale shows the magnitude of the out-of-plane component $\hat{S}_{z}$, and the arrows show the magnitude and direction of spins in the $k_{x}$-$k_{y}$ plane. (\textbf{b}) Simulated Hopf link with linking number one. The red (blue) circle represents the theoretical preimage of $\hat{\mathbf{S}}_{1}$ ($\hat{\mathbf{S}}_{2}$) and the scattered red squares (blue dots) are numerically simulated preimage of the $\epsilon$-neighborhood of $\hat{\mathbf{S}}_{1}$ ($\hat{\mathbf{S}}_{2})$ (see Methods section), which can be observed from time-of-flight images. The spin texture and preimages are computed by exactly diagonalizing the real space Hamiltonian with a lattice size $40\times40\times40$ under an open boundary condition. The parameters are chosen as $p=q=1$, $h=2$, $\gamma_{\text{t}}=0.01$, $\gamma_{\text{r}}=0.01$, and $\epsilon=0.15$. \label{fig:Spin-texture-and}}
\end{figure}

As discussed in the previous section, a nonvanishing $\chi(\hat{\mathbf{S}})$ implies a nontrivial spin texture. With cold atoms, one can actually visualize this nontrivial spin texture in the momentum space. In Fig.\ \ref{fig:Spin-texture-and}\textbf{a}, we plot a slice of the observed $\hat{\mathbf{S}}(\mathbf{k}$) with $k_{z}=0$. Although
this 2D plot cannot display full information of the 3D spin
texture, its twisted spin orientations do offer a glimpse of the whole nontrivial structure. It is worthwhile to note that this texture is distinct from typical 2D skyrmion \cite{skyrme1962unified} configurations, wherein swirling structure is a prominent feature \cite{yu2010real, rossler2006spontaneous}. More slices can be mapped out from the experiment to look for traits of a Hopfion.
With $\hat{\mathbf{S}}(\mathbf{k})$, one may also explicitly see knots and links in experiment. In the ideal case and continuum limit, the preimages of two different orientations of $\hat{\mathbf{S}}(\mathbf{k})$ should be linked with a linking number equal to the Hopf index.
Hence, if we map out these preimages, we would obtain a link. However, a real experiment always involves various kinds of noises. As
a result, the measured $\hat{\mathbf{S}}(\mathbf{k})$ cannot be perfectly
accurate. Moreover, due to the finite size, the observed $\hat{\mathbf{S}}(\mathbf{k})$
is discrete and only has a finite resolution. To circumvent these difficulties, we keep track of the preimages of a small neigborhood of the chosen orientations (see Methods section). We plotted a Hopf link in Fig.\ \ref{fig:Spin-texture-and}\textbf{b} based on the numerically simulated experimental results. From this figure and many other numerical simulations (not shown here for conciseness), we find that this link is stable against experimental imperfections. Varying $h$
and other parameters characterizing the strength of noise change
the shape of each circle, but the linked structure persists with linking
number one. \newline

\noindent \textbf{\large{Discussion}}

\noindent Hopf insulators are a special class of topological insulators beyond the periodic table for topological insulators and superconductors
\cite{schnyder2008classification, kitaev2009periodic}. Their physical
realization is of great importance but also especially challenging. With ultracold atoms in optical lattices, Hopf insulators could be realized using the Raman-assisted hopping technique \cite{jaksch2003creation,miyake2013realizing,aidelsburger2013realization,aidelsburger2011experimental}. The basic idea is analogous to that in Ref.\ \cite{wang2014probe}, where a cold-atom
implementation of 3D chiral topological insulators was proposed. Here we focus on how to measure the Hopf invariant and probe various intriguing knots and links, leaving a possible implementation protocol to the Supplementary Information. Other simpler experimental scenario for realizing Hopf insulators with cold atoms might also exist. A promising alternative is to consider periodically driven quantum systems, where modulation schemes can be tailored to implement diverse gauge fields \cite{goldman2014periodically} and an experimental realization of the Haldane model has been reported  \cite{Jotzu2014Experimental}.


Hopf insulators provide a new platform for exploring the deep connection and interplay between knot theory and topological phenomena, on which our knowledge still remain limited. For instance, in the above discussions, we showed that many different kinds of knots and links are hidden in Hopf insulators. However, a complete list of such knots and links is still lacking and requires further investigations. These studies will reveal new fundamental physical principle, and in the same time, can offer new guiding principles for experimental studies. From Fig.\ \ref{Hopflink}, it seems that partial information of certain spin orientations is already sufficient to characterize the topology of the Hopf insulators. This observation may help simplify the experimental detections. Taking the case in Fig.\ \ref{Hopflink}\textbf{e} as an example, to determine whether the system has a nontrivial topology or not, we only need to do the time-of-flight measurements in the $\hat{S}_y$ (or equivalently $\hat{S}_z$) basis. Thus, no rotation induced by fast Raman or radio frequency pulses is needed. In the future, it would be interesting to explore potential applications of the
knotted spin texture in designing spin devices based on topological states, which may have applications in spintronics and quantum information technologies \cite{Jungwirth2014Spin}.

In summary, we have shown that Hopf insulators support rich and highly nontrivial knot and link structures. We also demonstrated, via numerical simulations, that these exotic knots and associated nontrivial spin textures, which are robust to typical experimental imperfections, can be probed through time-of-flight imaging in cold atoms. Our results shed new light on the studies of topological insulators and open up a possibility to probe exotic knots and links in the cold-atom experimental platform. \newline

\small

\noindent \textbf{\large{Methods}}

\noindent \textbf{Stereographic coordinates.} For Hopf insulators in a cubic lattice, the first BZ is a 3D torus $\mathbb{T}^{3}$. Since
it is not convenient to draw and visualize different knots and links
in $\mathbb{T}^{3}$, we first do a map $g$ to go from $\mathbb{T}^{3}$
to $\mathbb{S}^{3}$ and use a stereographic coordinate system to
represent $\mathbb{S}^{3}$. The stereographic projection used in
this paper is defined as:
\begin{equation}
(x,y,z)  =  \frac{1}{1+\eta_{4}}(\eta_{1},\eta_{2},\eta_{3}),
\end{equation}
where $(x,y,z)$ and $(\eta_{1},\eta_{2},\eta_{3},\eta_{4})$ are
points of $\mathbb{R}^{3}$ and $\mathbb{S}^{3}$, respectively. \newline

\noindent \textbf{Perturbations to the Hamiltonian.}  In realistic experiments, there are additional noises other than the ideal Hamiltonian given by Eq.\ \eqref{eq:Hopf Ham-MomentumS}. The first one is a weak global harmonic trap typically present in cold-atom experiment. It is of the form $H_{\text{trap}}=\frac{1}{2}m\omega^{2}\sum_{\mathbf{r},\sigma}d_{\mathbf{r}}^{2}c_{\mathbf{r},\sigma}^{\dagger}c_{\mathbf{r},\sigma}$,
where $\sigma=\uparrow,\downarrow$, $m$ is the mass of the atom
and $d_{\mathbf{r}}$ is the distance from the center of the trap
to the lattice site $\mathbf{r}$. We use $\gamma_{\text{t}}=m\omega^{2}a^{2}/2$ to parametrize the relative strength of the trap with $a$ denoting
the lattice constant. The other perturbation we consider is a random noise  of the form $H_{\text{rand}}=\gamma_{\text{r}}\sum_{\mathbf{r},\mathbf{r}',\sigma,\sigma'}c_{\mathbf{r},\sigma}^{\dagger}\mathcal{R}_{\mathbf{r}\sigma,\mathbf{r}'\sigma'}c_{\mathbf{r}',\sigma'}$
where $\gamma_{\text{r}}$ characterizes the strength of the noise
and $\mathcal{R}$ is a random Hermitian matrix with its largest eigenvalue
normalized to unity. \newline

\noindent \textbf{Seeing links and knots from the time-of-flight data.} As discussed
in the main text, the spin orientation $\hat{\mathbf{S}}(\mathbf{k})$ in
real experiment is always pixelized with a finite resolution. Therefore,
for a specific orientation, $\hat{\mathbf{S}}_{1}$ for instance,
the measured $\hat{\mathbf{S}}(\mathbf{k})$ can only be approximately
rather than exactly equal to $\hat{\mathbf{S}}_{1}$ at any momentum point
$\mathbf{k}$. Consequently, one need to consider a small
$\epsilon$-neigborhood of $\hat{\mathbf{S}}_{1}$:
\begin{equation}
N_{\epsilon}(\hat{\mathbf{S}}_{1})  =  \{\hat{\mathbf{S}}:\;|\hat{\mathbf{S}}-\hat{\mathbf{S}}_{1}|\leq\epsilon\},
\end{equation}
where $|\hat{\mathbf{S}}-\hat{\mathbf{S}}_{1}|=[(\hat{S}_{x}-\hat{S}_{1x})^{2}+(\hat{S}_{y}-\hat{S}_{1y})^{2}+(\hat{S}_{z}-\hat{S}_{1z})^{2}]^{1/2}$
measures the distance between $\hat{\mathbf{S}}$ and $\hat{\mathbf{S}}_{1}$.
Let us denote the preimages of all orientations in $N_{\epsilon}(\hat{\mathbf{S}}_{1})$
as a set $P_{\epsilon}(\hat{\mathbf{S}}_{1})=(f\circ g)^{-1}[N_{\epsilon}(\hat{\mathbf{S}}_{1})]$.
With a finite resolution, the BZ is discrete and contains finite
momentum points, and so does $P_{\epsilon}(\hat{\mathbf{S}}_{1})$.
As a result, one should wisely choose an appropriate value for $\epsilon$
so that $P_{\epsilon}(\hat{\mathbf{S}}_{1})$ contains a proper amount
of momentum points to depict the loop structure of $(f\circ g)^{-1}(\hat{\mathbf{S}}_{1})$.
To obtain Fig.\ \ref{fig:Spin-texture-and}\textbf{b} in our numerical
simulation, we examine the discrete $\hat{\mathbf{S}}(\mathbf{k})$
(observed from time-of-flight measurements) at each
momentum point $\mathbf{k}$ and append $\mathbf{k}$ into the set
$P_{1\epsilon}$ ($P_{2\epsilon}$) if $\hat{\mathbf{S}}(\mathbf{k})$
is in a $\epsilon$-neigborhood of $\hat{\mathbf{S}}_{1}$ ($\hat{\mathbf{S}}_{2}$). Then Fig.\ \ref{fig:Spin-texture-and}\textbf{b} can be obtained by
plotting $g(P_{1\epsilon})$ and $g(P_{2\epsilon})$ in the stereographic
coordinate system defined above. \newline

\noindent \textbf{\large{Acknowledgement}}

\noindent We thank X.-J.\ Liu, G.\ Ortiz, J. E. Moore, H. Zhai, and D. Thurston for helpful discussions. D.L.D., S.T.W. and L.M.D. are supported by the ARL, the IARPA LogiQ program, and the AFOSR MURI program, and acknowledge support from Tsinghua University for their visits. K.S. acknowledges support from NSF under Grant No. PHY1402971. D.L.D. is also supported by JQI-NSF-PFC and LPS-MPO-CMTC at the final stage of this paper. \newline

\noindent \textbf{\large{Author contributions}}

\noindent L.M.D. supervised the project. All authors conceived the idea, analysed and discussed the results. D.L.D. performed the numerical simulations. All authors contributed to the writing of the manuscript.

\noindent \textbf{\large{Additional Information}}

\noindent \textbf{Supplementary Information} accompanies this paper on:

\noindent \textbf{Competing financial interests:} The authors declare no competing financial interests.

\bibliographystyle{naturemag}
\bibliography{Dengbib}

\clearpage
\onecolumngrid
\setcounter{figure}{0}
\makeatletter
\renewcommand{\thefigure}{S\@arabic\c@figure}
\setcounter{equation}{0} \makeatletter
\renewcommand \theequation{S\@arabic\c@equation}
\renewcommand \thetable{S\@arabic\c@table}

\begin{center} 
{\Large \bf Supplementary Information: Probe knots and Hopf insulators with ultracold
atoms}
\end{center}

In the main text, we have been focused on how to measure the Hopf invariant and probe knots and links using the stardard time-of-flight imaging techniques. In this Supplementary Information, we propose a possible experimental scheme to realize Hopf insulators with ultracold atoms in optical lattices. For simplicity, we consider the case of $p=q=1$.  After a Fourier transform, the corresponding Hamiltonian in real space reads
$H=H_{\text{sf}}+H_{\text{sp}}$, where $H_{\text{sf}}$ denotes the collection of all terms with spin flips:
\begin{eqnarray*}
H_{\text{sf}} & = & \sum_{\mathbf{r}}\frac{1}{2}[c_{\mathbf{r},\uparrow}^{\dagger}c_{\mathbf{r}+2\hat{x},\downarrow}-c_{\mathbf{r},\uparrow}^{\dagger}c_{\mathbf{r}-2\hat{x},\downarrow}-(1-i)c_{\mathbf{r},\uparrow}^{\dagger}c_{\mathbf{r}-\hat{x}-\hat{y},\downarrow}+(1+i)c_{\mathbf{r},\uparrow}^{\dagger}c_{\mathbf{r}+\hat{x}-\hat{y},\downarrow}-(1+i)c_{\mathbf{r},\uparrow}^{\dagger}c_{\mathbf{r}-\hat{x}+\hat{y},\downarrow}+ic_{\mathbf{r},\uparrow}^{\dagger}c_{\mathbf{r}-2\hat{y},\downarrow}\\
 &  & +(1-i)c_{\mathbf{r},\uparrow}^{\dagger}c_{\mathbf{r}+\hat{x}+\hat{y},\downarrow}-ic_{\mathbf{r},\uparrow}^{\dagger}c_{\mathbf{r}+2\hat{y},\downarrow}-2c_{\mathbf{r},\uparrow}^{\dagger}c_{\mathbf{r}-\hat{x}-\hat{z},\downarrow}+2c_{\mathbf{r},\uparrow}^{\dagger}c_{\mathbf{r}+\hat{x}-\hat{z},\downarrow}+2ic_{\mathbf{r},\uparrow}^{\dagger}c_{\mathbf{r}-\hat{y}-\hat{z},\downarrow}-2ic_{\mathbf{r},\uparrow}^{\dagger}c_{\mathbf{r}+\hat{y}-\hat{z},\downarrow}\\
 &  & +2h(c_{\mathbf{r},\uparrow}^{\dagger}c_{\mathbf{r}+\hat{x},\downarrow}-c_{\mathbf{r},\uparrow}^{\dagger}c_{\mathbf{r}-\hat{x},\downarrow}+ic_{\mathbf{r},\uparrow}^{\dagger}c_{\mathbf{r}-\hat{y},\downarrow}-ic_{\mathbf{r},\uparrow}^{\dagger}c_{\mathbf{r}+\hat{y},\downarrow})]+\text{h.c.},
\end{eqnarray*}
and $H_{\text{sp}}$ denotes the collection of all other terms:
\begin{eqnarray*}
H_{\text{sp}} & = & \sum_{\mathbf{r}}\frac{1}{2}[c_{\mathbf{r},\downarrow}^{\dagger}c_{\mathbf{r}-2\hat{x},\downarrow}+c_{\mathbf{r},\downarrow}^{\dagger}c_{\mathbf{r}+2\hat{x},\downarrow}+c_{\mathbf{r},\downarrow}^{\dagger}c_{\mathbf{r}+\hat{x}+\hat{y},\downarrow}+c_{\mathbf{r},\downarrow}^{\dagger}c_{\mathbf{r}-\hat{x}-\hat{y},\downarrow}+c_{\mathbf{r},\downarrow}^{\dagger}c_{\mathbf{r}-\hat{x}+\hat{y},\downarrow}+c_{\mathbf{r},\downarrow}^{\dagger}c_{\mathbf{r}+\hat{x}-\hat{y},\downarrow}+c_{\mathbf{r},\downarrow}^{\dagger}c_{\mathbf{r}-2\hat{y},\downarrow}\\
 &  & +c_{\mathbf{r},\downarrow}^{\dagger}c_{\mathbf{r}+2\hat{y},\downarrow}+c_{\mathbf{r},\downarrow}^{\dagger}c_{\mathbf{r}+\hat{x}+\hat{z},\downarrow}+c_{\mathbf{r},\downarrow}^{\dagger}c_{\mathbf{r}-\hat{x}-\hat{z},\downarrow}+c_{\mathbf{r},\downarrow}^{\dagger}c_{\mathbf{r}-\hat{x}+\hat{z},\downarrow}+c_{\mathbf{r},\downarrow}^{\dagger}c_{\mathbf{r}+\hat{x}-\hat{z},\downarrow}+c_{\mathbf{r},\downarrow}^{\dagger}c_{\mathbf{r}+\hat{y}+\hat{z},\downarrow}+c_{\mathbf{r},\downarrow}^{\dagger}c_{\mathbf{r}-\hat{y}-\hat{z},\downarrow}\\
 &  & +c_{\mathbf{r},\downarrow}^{\dagger}c_{\mathbf{r}-\hat{y}+\hat{z},\downarrow}+c_{\mathbf{r},\downarrow}^{\dagger}c_{\mathbf{r}+\hat{y}-\hat{z},\downarrow}+2h(c_{\mathbf{r},\downarrow}^{\dagger}c_{\mathbf{r}+\hat{x},\downarrow}+c_{\mathbf{r},\downarrow}^{\dagger}c_{\mathbf{r}-\hat{x},\downarrow}+c_{\mathbf{r},\downarrow}^{\dagger}c_{\mathbf{r}+\hat{y},\downarrow}+c_{\mathbf{r},\downarrow}^{\dagger}c_{\mathbf{r}-\hat{y},\downarrow}\\
 &  & +c_{\mathbf{r},\downarrow}^{\dagger}c_{\mathbf{r}+\hat{z},\downarrow}+c_{\mathbf{r},\downarrow}^{\dagger}c_{\mathbf{r}-\hat{z},\downarrow})+2(1+h^{2})c_{\mathbf{r},\downarrow}^{\dagger}c_{\mathbf{r},\downarrow}]-[\downarrow\rightarrow\uparrow].
\end{eqnarray*}

\begin{figure}[!t]
\includegraphics[width=0.68\textwidth]{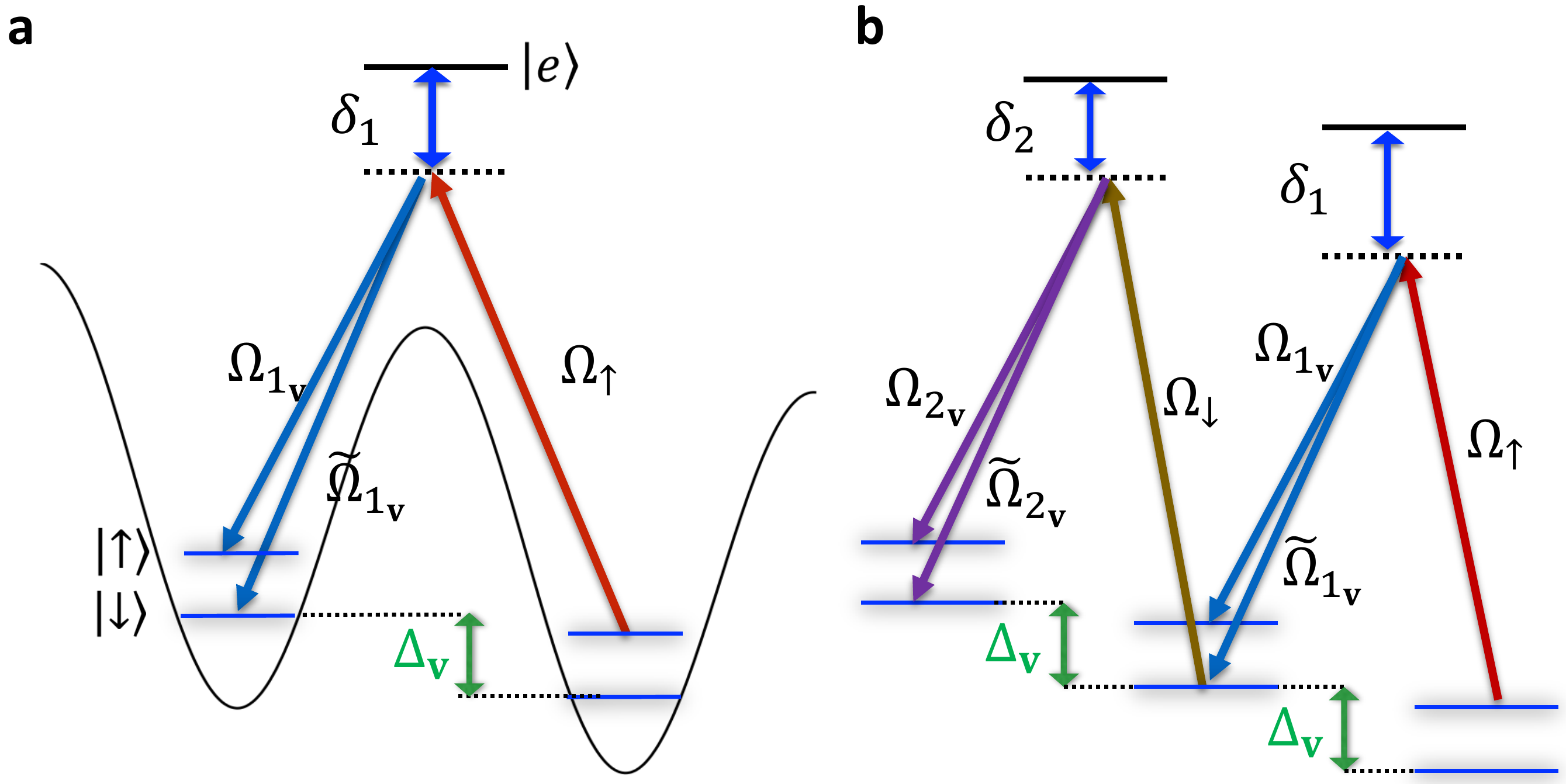}
\caption{\textbf{Laser configurations.} Schematics of the laser configuration to realize the Hamiltonian $H_{\mathbf{v}}$. (\textbf{a}) A linear tilt $\Delta_{\mathbf{v}}$ per site along the $\mathbf{v}$ direction
is imposed. A Raman triplet with detunings matching the frequency offset
can induce spin-flip hoppings in a desired direction. (\textbf{b})
Two Raman triplets that realize all hoppings in  $H_{\mathbf{v}}$. $\delta_{1}$ and $\delta_{2}$ are two different detunings from the excited states. The Rabi frequencies for each beam in terms of the unit $\Omega_{0}$ are: $\widetilde{\Omega}_{1\mathbf{v}}=-\sqrt{2}\Omega_{0}e^{i\pi/4}e^{-ikx}$,
$\Omega_{1\mathbf{v}}=-\Omega_{0}e^{-ikx}$, $\Omega_{\uparrow}=\Omega_{0}e^{iky}$,
$\widetilde{\Omega}_{2\mathbf{v}}=\sqrt{2}\Omega_{0}e^{-i\pi/4}e^{-ikx}$,
$\Omega_{2\mathbf{v}}=\Omega_{0}e^{-ikx}$, and $\Omega_{\downarrow}=\Omega_{0}e^{iky}$,
where $k$ is the magnitude of the laser wave vector. All these Raman
beams can be draw from a single laser through an electric or acoustic
optical modulator. \label{fig:Laser Configuration}}
\end{figure}

To realize Hopf insulators with ultracold atoms, the method used here is similar to that in Ref.~\cite{wang2014probe}, where a cold-atom implementation of three-dimensional (3D) chiral topological insulators was proposed. We consider cold fermionic atoms ($^{6}\text{Li}$ for example) in a 3D cubic optical lattice and choose two internal atomic levels (hyperfine states) as our spin states $\left| \uparrow \rangle \right.$ and $\left| \downarrow\rangle \right.$. Other levels are initially depopulated by
optical pumping and transitions to those levels are forbidden due
to a large energy detuning or carefully selected laser polarizations.
In real space, the Hamiltonian 
 has spin-orbit coupled hoppings along nine possible directions. Here, we explicitly demonstrate how to engineer the specific hoppings along the $\mathbf{v}=\hat{x}+\hat{y}$ direction. For other directions, the realization scheme will be similar and thus omitted for conciseness. The hopping terms in the $\mathbf{v}$ direction can be written as
\begin{align}
H_{\mathbf{v}}  =  \frac{1}{2}\sum_{\mathbf{r}}[-(1+i)c_{\mathbf{r},\downarrow}^{\dagger}-c_{\mathbf{r},\uparrow}^{\dagger}]c_{\mathbf{r}+\mathbf{v,}\uparrow} 
 + [(1-i)c_{\mathbf{r},\uparrow}^{\dagger}+c_{\mathbf{r},\downarrow}^{\dagger}]c_{\mathbf{r}+\mathbf{v,}\downarrow}+\text{h.c}.
\end{align}
Apparently, $H_{\mathbf{v}}$ consists of various hopping terms coupled with spin rotations. Basically, both spin $\left| \uparrow \rangle \right.$ and $\left| \downarrow\rangle \right.$ hop along the $-\mathbf{v}$ direction to become a superposition of both spin states. We can decompose the four different hopping terms and each of them can be achieved via Raman-assisted tunnelings \cite{jaksch2003creation, miyake2013realizing, aidelsburger2013realization}. For instance, the first two hoppings can be activated by two Raman pairs, $\Omega_{\uparrow}$, $\widetilde{\Omega}_{1\mathbf{v}}$, and $\Omega_{\uparrow}$, $\Omega_{1\mathbf{v}}$, where $\Omega_{\uparrow}$ is in common; $\widetilde{\Omega}_{1\mathbf{v}}$ and $\Omega_{1\mathbf{v}}$ can be drawn from the same beam split by an electric or acoustic optical modulator. The phase and strength of the hopping can be controlled by the laser phase and intensity (see caption of Fig.\ \ref{fig:Laser Configuration}). Similarly, the other two hopping terms can be triggered by the Raman triplet $\widetilde{\Omega}_{2\mathbf{v}}$, $\Omega_{2\mathbf{v}}$ and $\Omega_{\downarrow}$ as shown in Fig.\ \ref{fig:Laser Configuration}.

One may notice that the parity (left-right) symmetry is explicitly broken  and natural hoppings are suppressed. Both of these can be achieved by a homogeneous energy gradient along the $\mathbf{v}$ direction, which can be accomplished, for instance, by the magnetic field gradient,  dc- or ac-Stark shift gradient, or the natural gravitational field \cite{wang2014probe,miyake2013realizing,aidelsburger2013realization}. We denote the linear energy shift per site in the $\mathbf{v}$ direction by $\Delta_{\mathbf{v}}$ and impose that natural tunneling rate $t_{0}\ll\Delta_{\mathbf{v}}$. As a consequence, the natural tunneling probability $(t_{0}/ \Delta_{\mathbf{v}})^{2}$ is negligible in this tilted lattice. The energy offset also forbids Raman-assisted tunnelings in the opposite direction other than the ones prescribed above. Incommensurate tilts along different directions also suppress unwanted couplings among them. The detuning $\delta_{1}$ and $\delta_{2}$ for the two Raman triplets should also be different, so that no unintended interference between these two triplets happens. In addition, the wave-vector difference of two Raman beams $\delta \mathbf k$ has to have a component along the hopping direction ($\mathbf{v}$-direction in this case) to ensure the Raman-assisted hopping strength is non-vanishing \cite{wang2014probe}.

Since the real-space Hamiltonian contains hopping terms along nine different directions and each of them requires a similar configuration as above, the overall scheme necessitates a considerable number of laser beams. Although many of the beams can be used in common and created from the same laser by an electric or acoustic optical modulator, the implementation is nonetheless very challenging. A simpler implementation protocol of Hopf insulators is highly desired. Controlled periodic driving may be a promising alternative. Also, theoretically it may be possible to find a more natural model Hamiltonian that encapsulates the nontrivial Hopf physics, which may in turn simplify the experimental realization. We would like to emphasize, however, our detection and measurement protocol do not depend on the particular realization scheme. Regardless of the specific implementation of Hopf Hamiltonians, we can detect the Hopf invariant and probe the rich knot structures hidden in Hopf insulators. 

\end{document}